# An Open-Path Eddy-Covariance Laser Spectrometer for Simultaneous Monitoring of $CO_2$, $CH_4$, and $H_2O$

Viacheslav Meshcherinov[1,2]*, Iskander Gazizov[3], Bogdan Pravuk[1,2], Viktor Kazakov[1,2], Sergei Zenevich[1], Maxim Spiridonov[2], Shamil Gazizov[1], Gennady Suvorov[4,5], Olga Kuricheva[5], Yuri Lebedev[2], Imant Vinogradov[2] and Alexander Rodin[1]

[1] *Moscow Center for Advanced Studies, 123592 Moscow, Russia*

[2] *Space Research Institute of the Russian Academy of Sciences (IKI RAS), 117997 Moscow, Russia*

[3] *Institute of Chemical Technologies and Analytics, TU Wien, 1060 Vienna, Austria*

[4] *Institute of Forest Science of the Russian Academy of Sciences, 143030 Uspenskoe, Russia*

[5] *A.N. Severtsov Institute of Ecology and Evolution of the Russian Academy of Sciences, 119071 Moscow, Russia*

*Corresponding author. Email: viacheslav.meshcherinov@gmail.com

Contributing authors: iskander.gazizov@proton.me; pravuk.br@phystech.edu; kazakov@cosmos.ru; zenevich.sg@phystech.edu; maxim.spiridonov@gmail.com; shamil.gazizov@proton.me; suvorovg@gmail.com; olga.alek.de@gmail.com; yv.lebedev@yandex.ru; imant@cosmos.ru; alexander.rodin@phystech.edu

**Abstract**

We present E-CAHORS – a compact mid-infrared open-path diode-laser spectrometer designed for the simultaneous measurement of carbon dioxide, methane, and water vapor concentrations in the near-surface atmospheric layer. These measurements, combined with simultaneous data from a three-dimensional anemometer, can be used to determine fluxes using the eddy-covariance method. The instrument utilizes two interband cascade lasers operating at 2.78 μm and 3.24 μm within a novel four-pass M-shaped optical cell, which provides high signal power and long-term field operation without requiring active air sampling. Two detection techniques – tunable diode laser absorption spectroscopy (TDLAS) and a simplified wavelength modulation spectroscopy (sWMS) – were implemented and evaluated. Laboratory calibration demonstrated linear responses for all gases ($R^2 \approx 0.999$) and detection precisions at 10 Hz of 311 ppb for $CO_2$, 8.87 ppb for $CH_4$, and 788 ppb for $H_2O$. Field tests conducted at a grassland site near Moscow showed strong correlations (R = 0.91 for $CO_2$ and $H_2O$, R = 0.67 for $CH_4$) with commercial LI-COR LI-7200 and LI-7700 analyzers. The TDLAS mode demonstrated lower noise and greater stability under outdoor conditions, while sWMS provided baseline-free spectra but was more sensitive to power fluctuations. E-CAHORS combines high precision, multi-species sensing capability with low power consumption (10 W) and a compact design (4.2 kg).

## 1. Introduction

People have long suspected that human activities can alter the local climate. As early as Theophrastus (c. 371 – c. 287 BC), scholars debated whether deforestation and land drainage could alter temperature and rainfall around Greek cities such as Larissa and Philippi [1]. However, the scientific study of climate change began in earnest only in the 19th century, following the discovery of the natural greenhouse effect [2–5]. Despite subsequent advances, many remote regions still lack precise, on-site measurement stations. This paper will focus on developing a portable instrument to measure natural abundance of greenhouse gases in local ecosystems, with the ability to retrieve fluxes with the Eddy Covariance (EC) method.

Over the past 30 years, the EC method has been extensively applied in studies of the Earth's atmospheric boundary layer to measure vertical turbulent heat fluxes, water vapor, carbon dioxide, and other gases [6]. This method directly measures greenhouse gas fluxes between vegetated land surfaces and the atmosphere. Modern commercial instruments suitable for this method enable the measurement of mass and energy fluxes at hourly to annual scales, with minimal impact on local ecosystems. Typically, the surface area covered by a single measurement station ranges from about 100 to 2000 meters [7]. Thus, deploying EC stations as part of a coordinated network can provide quantitative assessments of how large-scale ecosystems respond to human-induced anthropogenic influences.

The concept of flux can be defined as the amount of a substance passing through a closed surface per unit of time. Fluxes are derived from gas concentration measurements and wind velocity vectors. Air movement above the surface can be visualized as horizontal transport by numerous rotating eddies. The wind speed within each eddy has a vertical component, which is measured by an ultrasonic anemometer. Simultaneously, the concentration of the target gas is measured by a spectrometer. Thus, the fundamental principle of the EC method is based on calculating the covariance between the concentration of the studied gas and the vertical wind velocity within air eddies [8]. In the eddy-covariance framework, the turbulent vertical flux $F$ [g/(m$^2$s)] of a trace gas is therefore written as [8]:

$$F \approx \overline{\rho_d}\, \overline{w's'}. \tag{1}$$

This expression is obtained by decomposing vertical wind speed and gas concentration into mean and fluctuating components and retaining the covariance term, while assuming that density fluctuations and the mean vertical flow are negligible. Here $\rho_d$ [g/m$^3$] is the mean dry air density, $w'$ [m/s] is the instantaneous deviation of vertical wind velocity from its mean value, and $s'$ [ppm] is the corresponding deviation of the gas dry mole fraction (or mixing ratio). The overbar denotes averaging over a suitable time interval, so that $\overline{w's'}$ represents the covariance between velocity and concentration fluctuations that drives the vertical turbulent flux.

*1.1. State of the Art*

Various gas analyzers are available to measure gas concentrations at frequencies suitable for EC. These analyzers can be categorized based on the air's pathway before reaching the measurement chamber: open-path, closed-path, and enclosed-path sensors. In an open-path system, the analyzer is directly exposed to the surrounding air, allowing gases to be measured with minimal disturbance. In contrast, a closed-path system draws air through a tube into a controlled sampling cell within the analyzer, which allows operating under rain or snow, but might restrict high-frequency eddies and requires an additional pump [9–11]. Comparisons of different types of such sensors have shown promising agreement [12,13].

Commercial gas analyzers can also be classified based on their operating principles: absolute Non-Dispersive Infrared (NDIR) analyzers, which are characterized by a broad spectral range, and diode laser-based analyzers, which offer high spectral resolution but operate over a narrow spectral range. The advantages of laser-based analyzers include high spectral selectivity and the capability to detect complex molecules. In contrast, NDIR analyzers benefit from measuring multiple absorption lines or even entire absorption bands, which helps eliminate temperature dependence and significantly reduces sensitivity to pressure variations [14].

One of the popular commercial instruments is the LI-7200 (LI-COR Inc., USA) – a compact closed-path $CO_2/H_2O$ analyzer based on the NDIR design of the open-path LI-7500, adapted for short intake tubing [15]. It is often combined with a laser-based LI-7700 open-path methane analyzer (LI-COR), which employs a Herriott cell (0.47 m base, 30 m optical path) [16] and determines methane concentration via wavelength modulation spectroscopy (WMS) near 1.65 μm at 20 Hz [17]. The signal is demodulated at twice the modulation frequency and normalized to laser intensity [18].

To achieve faster measurements, a 100 Hz gas analyzer using second-derivative laser absorption spectroscopy was developed, holding significant potential for flux measurement applications, particularly in environments characterized by rapid turbulence [19]. Another group developed a dual-laser WMS instrument for $CO_2$ and $H_2O$ fluxes featuring an anti-pollution optical cell combining a Herriott multipass and a dual-pass design [20]. Notably, the IRGASON instrument (Campbell Scientific Inc., USA) integrates an ultrasonic anemometer with an NDIR analyzer in one design, instead of having multiple separate devices [21].

Beyond the near-infrared, several mid-infrared analyzers have been introduced based on quantum cascade lasers, such as the open-path HT8600 (3.22 μm, $CH_4$) and HT8700 (9.06 μm, $NH_3$) from Healthy Photon Co., China, suitable for tower installation, though with performance not exceeding the competitors [22,23]. Closed-path systems offering higher sensitivity but far less field flexibility include the FMA (Fast Methane Analyzer) from Los Gatos Research, USA [24], QCLAS (Aerodyne Research, USA) [25], and G2311-f (Picarro, USA) [26]. These instruments operate around 10 Hz and require multi-metre inlet tubes, making them less practical for tower deployment.

This paper describes the development, calibration, and field testing of Eddy-Covariance Atmospheric fluxes High-speed Optical Research Spectrometer (E-CAHORS), a novel instrument designed to monitor three primary greenhouse gases via eddy covariance measurements. Our goal is to develop a reliable, commercially viable sensor using a simple yet effective design. By employing a four-pass open-path optical system, the instrument maintains a high signal power over long periods of field operation, even when dust accumulates on the windows. The open-path configuration removes the need for air pumps, simplifying installation on tall towers. Choosing a laser-based approach over traditional NDIR sensors enables flexibility in targeting different analytes in the future. Additionally, the use of increasingly affordable mid-infrared components helps simplify the overall design, removing the need for a multipass cell. Finally, we introduce a cross-pattern M-shaped optical scheme that allows for the simultaneous measurement of $CO_2$, $CH_4$, and $H_2O$ using two lasers.

## 2. Materials and Methods

### 2.1. Spectral Range

The methane absorption lines near 3.24 μm (3086 cm$^{-1}$), corresponding to the fundamental $v_3$ mode of C-H bond vibrations, are approximately 40 times more intense than the overtone lines of this mode near 1.65 μm (6057 cm$^{-1}$). For this reason, the 3.24 μm spectral range was selected for the $CH_4$ detection channel in the design of the E-CAHORS prototype. This choice enabled the use of a relatively simple four-pass optical configuration with a comparatively short optical path, which is easier to design and align than multi-pass optical systems, such as Herriott or Chernin cells [16,27,28]. Figure 1(a) depicts the simulation of this spectral line based on the Voigt profile [29] parameters from the HITRAN-2020 database [30]. This line, free from overlap with $H_2O$ absorption lines, remains detectable even under moderate to high humidity conditions.

Selecting an appropriate spectral range for analyzing carbon dioxide and water vapor concentrations presents a more complex challenge. The 4.2 μm range, which includes strong $CO_2$ absorption lines, presents challenges in sourcing high-quality lasers and photodiodes at prices suitable for commercial instruments. Conversely, the 2.0 μm range offers suitable diode lasers and extended InGaAs-photodiodes for laser spectroscopy applications. However, the $CO_2$ absorption in this region is not sufficiently intense for the chosen optical design, which features a relatively short optical path. Consequently, the 2.8 μm range was selected, as it provides $CO_2$ absorption of adequate intensity. However, this range also contains strong water vapor absorption lines. To mitigate this issue, we chose a spectral region that includes a weak water vapor absorption relative to its neighboring lines and two $CO_2$ absorption lines, which is expected to enhance the accuracy of $CO_2$ concentration retrieval. Nevertheless, the ICLs could be easily switched for different spectral ranges without any modifications to the instrument.

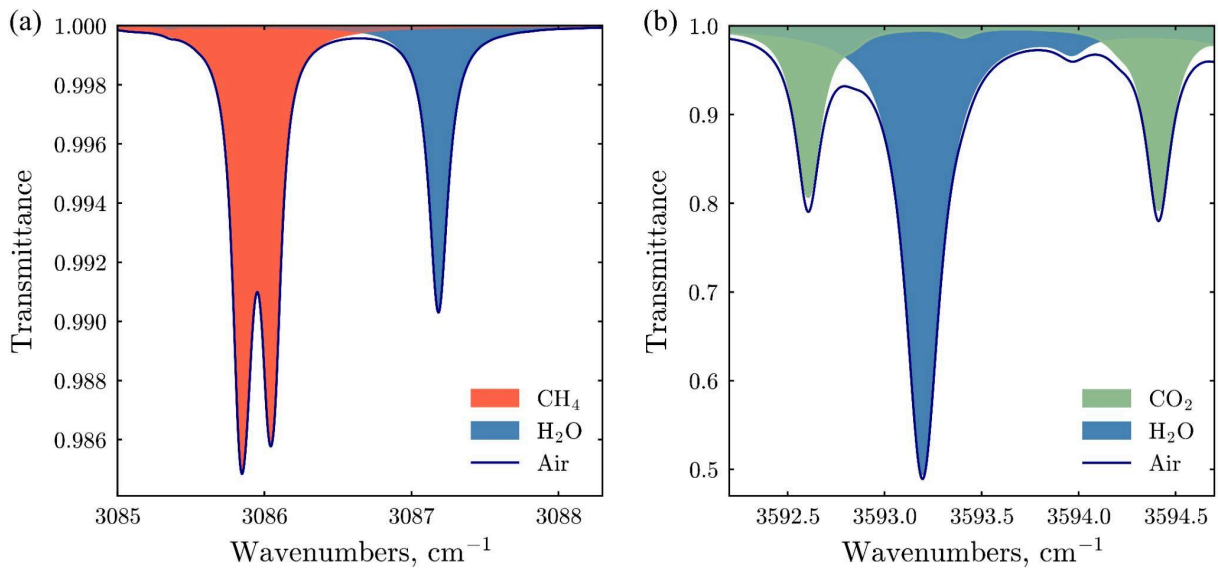

**Figure 1.** Spectral transmittance of air (2.2 ppm $CH_4$, 3000 ppm $H_2O$, 430 ppm $CO_2$) within **(a)** 3085-3088 cm$^{-1}$ range and **(b)** 3592-3595 cm$^{-1}$; at a temperature of 298 K, pressure of 1 atm, and a distance of 140 cm, derived from the HITRAN-2020 database [30].

*2.2. Absorption Spectroscopy*

Laser diodes combine high-intensity stability with rapid frequency tuning. However, as absorption spectroscopy techniques evolved, it became evident that acquiring information at higher frequencies provides a more suitable approach for detecting weaker signals due to 1/*f* noise. Thus, modulation techniques such as WMS [18] and frequency modulation spectroscopy (FMS) [31,32] have gained popularity.

Yet, comparison studies demonstrate that even traditional tunable diode laser absorption spectroscopy (TDLAS) yields similar performance if the laser is swept quickly [33,34]. It should be noted that the widely used TDLAS technique offers advantages in specific applications, including isotopic analysis, the principles of which were thoroughly explored in our previous works [35,36]. On the other hand, modulation techniques address baseline-related issues, as the demodulated signal has no DC baseline [37,38]. Thus, in this work, we utilize both the direct absorption technique and one of the WMS technique options to determine which one fits the Eddy Covariance application better.

*2.3. Basic Principles of the sWMS Technique*

In the case of classical WMS, the laser current is modulated simultaneously by a low-frequency sawtooth signal and a higher-frequency sinusoidal signal [20]. The sawtooth modulation enables slow spectral scanning near one or multiple absorption lines to analyze their shape, with typical frequencies ranging from a few to several tens of hertz. Meanwhile, the sinusoidal modulation of the laser current, usually at a ~kHz frequency, induces additional frequency and intensity modulation of the laser [39,40]. The photodiode signal is then demodulated at the harmonics of the modulation frequency $f$ with a lock-in amplifier to construct a spectrum. A simplified approach to WMS also exists (sWMS), which does not require a lock-in amplifier [41]. This method is based on applying a rectangular modulation scheme, which is sufficient for calculating the second quasi-derivative of the detected signal.

Modulation in sWMS is achieved by periodically modulating the injection current in a triangular shape, as shown in Figure 2(a). Each period consists of four points. Points 2 and 4 maintain the same current level, resulting in similar laser emission intensities, while points 1 and 3 are related to the lower and higher current values, respectively. However, the laser emission frequencies at points 2 and 4 differ. This discrepancy arises because of the differences in laser temperature caused by the previous modulation points, as mentioned in [41]. The acquired signal, presented in Figure 2(b), is divided into four distinct data arrays, as shown in Figure 2(c). The radiation intensities observed in the arrays $I_2$ and $I_4$ corresponding to points 2 and 4 of the modulation exhibit near-identical values. Arrays $I_1$ and $I_3$ have the widest frequency shifts. From these arrays, one could calculate the first derivative $I_{d1}$ and the second derivative $I_{d2}$ of the signal, as presented by the equations:

$$I_{d1} = I_3 - I_1, \quad I_{d2} = I_3 + I_1 - 2I_2. \tag{2}$$

Analysis of the frequency differences between the first and third modulation arrays aligns with conventional modulation techniques. In contrast, the second and fourth arrays are associated with a modulation method incorporating non-stationary heating and cooling of the laser's active region [41]. As illustrated in Figure 2(d), the second derivative of the signal $I_{d2}$ exhibits no offset, whereas the first derivative shows a noticeable offset. This makes the second derivative suitable for analyzing the concentration of the target gas component in ambient air – after calculating the second derivative, the spectrum is divided into segments corresponding to individual atmospheric absorption lines, and the integral under each line is calculated. This integral represents the raw, uncalibrated spectrometer output, which is proportional to the true gas concentration. To obtain quantitative results, a standard calibration procedure is then applied using reference gases with known concentrations.

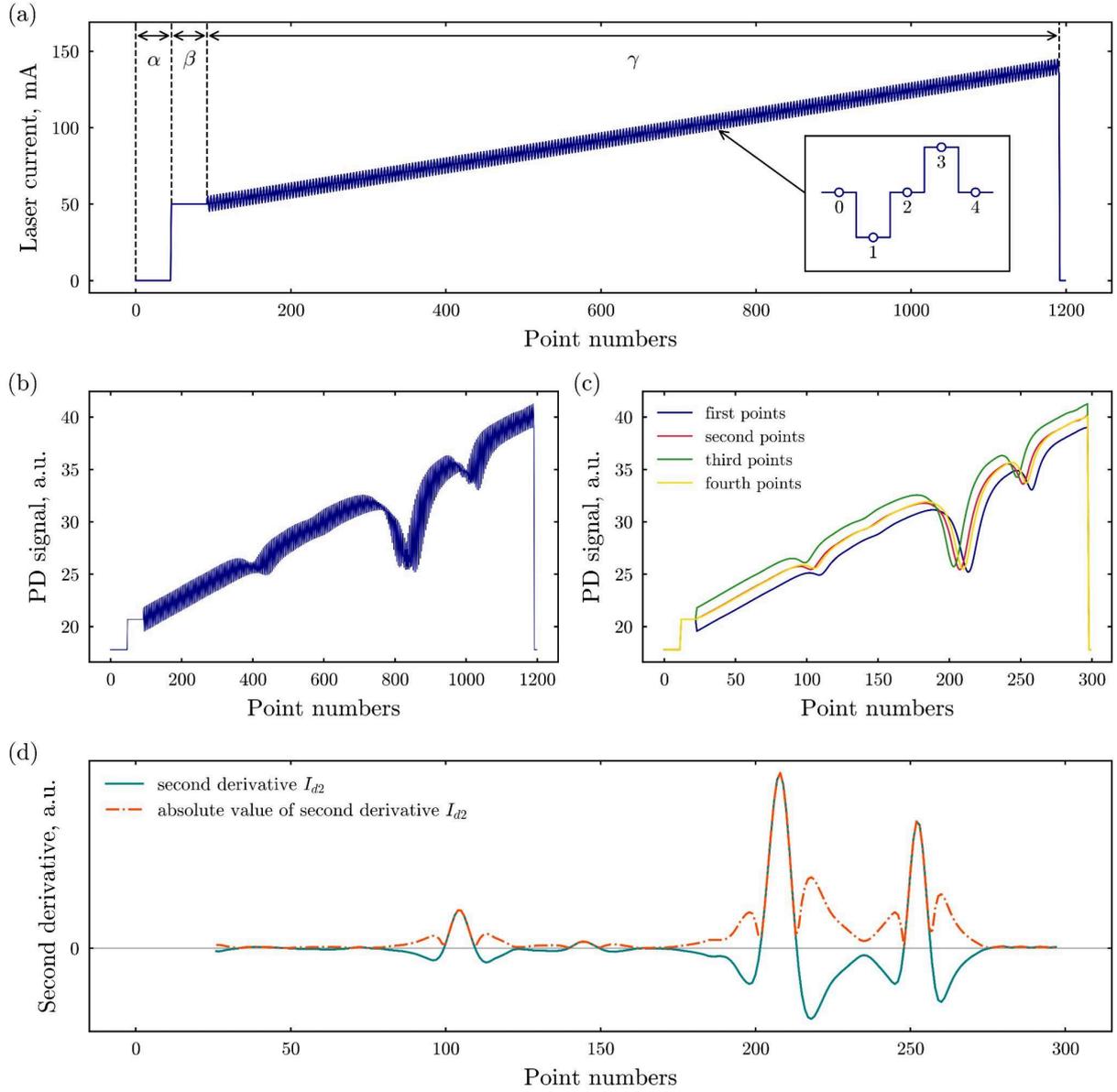

**Figure 2. (a)** Laser injection current form for a single period, the inset illustrates the shape of one modulation cycle of the injection current; **(b)** detected signal; **(c)** detected signal partitioned into arrays of corresponding data points (right); **(d)** the second derivative $I_{d2}$ of the detected signal and the absolute value of $I_{d2}$. The injection current cycle consists of three distinct phases: (*α*) a no-pumping region, (*β*) a steady-state laser emission region, and (*γ*) a region of spectral scanning and laser frequency modulation.

## 2.4. TDLAS Spectra Processing

In TDLAS, the laser light sweeps across the absorption feature, and the photodiode detects intensity $I(v)$, which is the convolution of the transmission spectrum $T(v)$ and laser intensity modulation $I_0(v)$ due to the current ramp, where $v$ is the corresponding wavenumber value. An advantage of TDLAS is that the calculated absorption coefficient $\alpha(v)$ [cm$^{-1}$] spectrum can be directly fitted using spectral line parameters, which results in calibration-free quantification of target gases. The basic principle of TDLAS is therefore described by the Bouguer-Lambert-Beer law:

$$\frac{I(\nu)}{I_0(\nu)} = T(\nu) = e^{-\alpha(\nu)L} = e^{-N\sigma(\nu)L}, \qquad (3)$$

where $\sigma$ [cm$^2$/molecule] is the absorption cross section, $L$ [cm] is the optical path length, and $N$ [molecules/cm$^3$] is the number density of absorbing species. Extracting the transmission spectrum is challenging and requires an accurate estimation of the laser intensity modulation $I_0$. To retrieve the absorption spectrum, we apply Gram-Schmidt orthogonalization to generate a set of orthogonal polynomials that capture the low-frequency trends in the original spectrum, allowing for a reliable estimation of $I_0$ [42,43]. The orthogonalization is then applied to both the model and experimental data, thus eliminating the need for baseline estimation. For the gas model, we calculate the Voigt profiles with the HITRAN-2020 database and the hapi Python library [30,44]. Finally, the number density $N$ is retrieved by fitting the measured spectra to the model using non-linear least-squares regression, which enables the calculation of instantaneous partial pressure values $p_{self}$ [atm]:

$$N = N_L \left( \frac{T_0}{T_{air}} \frac{p_{self}}{p_0} \right), \qquad (4)$$

where $N_L$ is the Loschmidt constant [molecules/cm$^3$], $T_{air}$ is an instantaneous value of temperature [K], $T_0$ = 273.15 K, and $p_0$ = 1 atm. The gas species mixing ratio could be then calculated as $p_{self}/p_{air}$.

*2.5. Instrument Design*

The instrument design was guided by an open-cell configuration concept, enabling *in situ* operation without the need for a pump to circulate the gas sample through the analytical volume. This approach contributed to the instrument's compactness and low mass, simplifying its installation on towers tens of meters high, and ensured minimal disruption to incoming air fluxes. A four-pass optical system, combined with approximately 5 mW lasers, allows for a reliable signal detection even when dust accumulates on optical windows, thereby extending the maintenance-free operation time. The standalone instrument and an illustration of two laser beam paths within the open cell are shown in Figure 3.

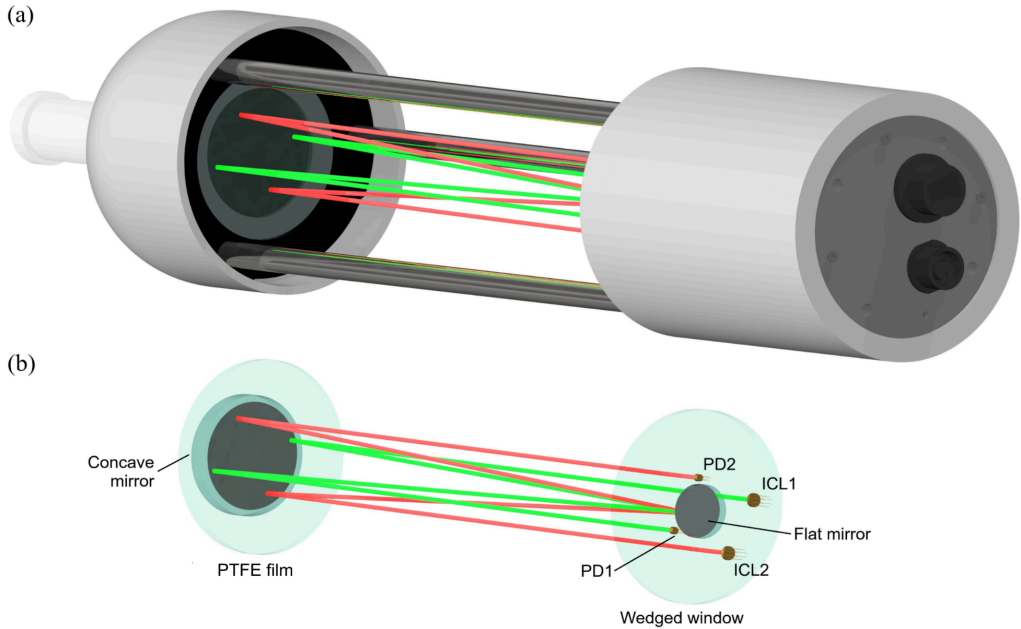

**Figure 3. (a)** The E-CAHORS prototype. **(b)** Prototype optical scheme: ICL – interband cascade laser, PD – photodiode.

### 2.4.1. Optics

The instrument's optical design is based on a four-pass open cell with laser beam paths arranged in an X-shaped configuration. The beam paths of the two lasers are depicted in red and green lines, respectively, in Figure 3(b). We chose two interband cascade lasers (ICL) operating at wavelengths of 3.24 μm and 2.78 μm as the light sources (LD-PD, Singapore). Each laser is equipped with a collimating lens C036TME-D (Thorlabs, USA).

Each of the two laser beams passes through a sapphire wedged window, which protects the main instrument housing, containing the control electronics and photonic components, from environmental exposure. The beam then traverses the atmospheric air for the first time, enters through the PTFE film, and is reflected by a spherical mirror with a 350 mm focal length, made of fused silica with an aluminum coating (R > 95%). After passing through the atmospheric air again, it is reflected by a flat mirror made of fused silica with an aluminum coating. Following an additional reflection from the spherical mirror, the beam completes its M-shaped optical path and reaches the PD34 photodiode (IoffeLED, Russia). The total optical path length for each laser beam is 135 cm.

### 2.4.2. Electronics

The instrument is designed for outdoor use and can operate in rain, snow, and within a temperature range of -20 to 40℃. Its thermal management allows the direct transfer of heat from electronic modules to the instrument housing, ensuring heating during winter and cooling of electronic components during summer. A schematic diagram of the instrument is shown in Figure 4, and the specifications are detailed in Table 1.

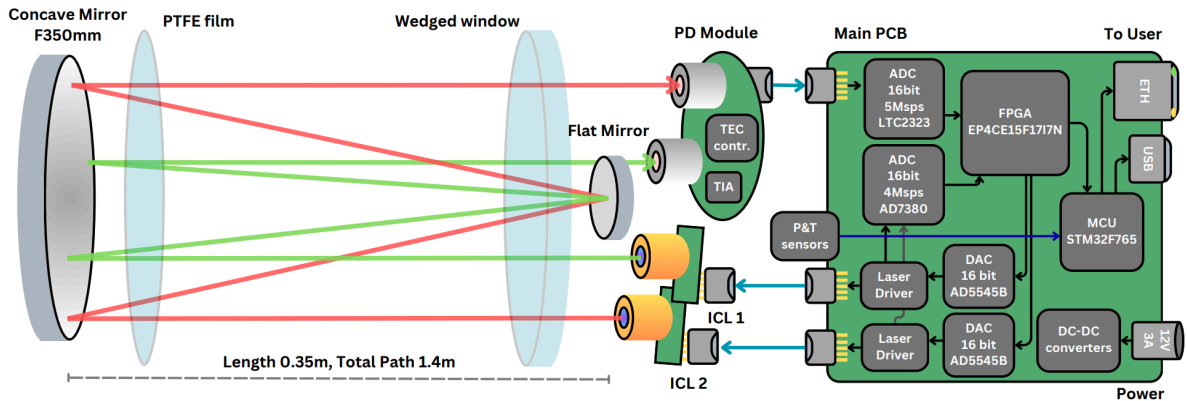

**Figure 4.** Block diagram of the electric circuits and principal optical diagram of the instrument.

The instrument's electronics consist of three modules: a main board, a photodiode board, and two laser boards. The main printed circuit board (PCB), measuring 85 mm x 85 mm, controls two lasers simultaneously, reads data from two photodiodes, averages the signals, and transmits the data to a PC. An EP4CE15F17I7N FPGA (Intel, USA) was selected to control two spectral channels. This FPGA manages two dual-channel 16-bit AD5545BRUZ DACs (Analog Devices, USA) with a sampling frequency of 2 MHz, for controlling laser Peltier modules and setting laser diode currents. Additionally, the FPGA reads data from a dual-channel 16-bit AD7380 ADC (Analog Devices, USA) at a frequency of 4 MHz for digitizing photodiode signals, and from two dual-channel 16-bit LTC2323 ADCs (Analog Devices, USA) for digitizing internal laser thermistors and external instrument thermistors to measure air temperature. Averaged spectra are then transmitted to an STM32F765NIH6 microcontroller (STMicroelectronics, Switzerland) for PC communication via Ethernet or USB 2.0. The microcontroller also decodes packets from the PC to adjust laser parameters. We use DC-DC converters to power the digital electronics and Low-dropout regulators (LDO) for the analog components, thereby improving the signal-to-noise ratio (SNR). Power to the instrument is supplied at 8.5 V (10 W).

The photodiode board consists of two Mid-IR photodiodes. Signals from these photodiodes are amplified and filtered using operational amplifiers, converted into differential signals, and transmitted to the main board. The photodiode board also features an analog proportional-integral controller that automatically cools the photodiodes to -30 °C. The heat from the photodiodes is dissipated via an aluminum plate located between the board and the photodiodes. The plate is mounted to the main assembly to allow for heat transfer to the instrument's housing.

To monitor the real-time pressure and temperature of the gas mixture inside the vacuum chamber during the E-CAHORS instrument calibration, a quartz pressure transducer PDA-0.106-0.025 (SKTB ElPA, Russia) with a measurement accuracy of 0.25 mbar and a quartz temperature transducer PTK-5-200-03 (SKTB ElPA, Russia) with an accuracy of 0.03°C were employed. During the measurements of greenhouse gas fluxes in field conditions, the same pressure and temperature sensors were used.

**Table 1.** Specification of the Instrument. LOD is the limit of detection, determined from Allan-Werle analysis at the longest stable integration time.

| LOD, 1$\sigma$ | Wavelengths | Mass | Size | Optical Power | Sample rate | Power draw | Operating temperature |
|---|---|---|---|---|---|---|---|
| 57.5 ppb ($CO_2$) 0.3 ppb ($CH_4$) 129 ppb ($H_2O$) | 2782 nm ($CO_2$, $H_2O$) 3241 nm ($CH_4$) | 4.2 kg | ⌀130 mm, 670 mm height | 5.5 mW ($CO_2$, $H_2O$) 8.5 mW ($CH_4$) | up to 20 Hz | 10 W | -20℃ to 40℃ |

## 3. Results and Discussion

During the tests of the E-CAHORS prototype, the instrument was calibrated using known gas mixtures in a vacuum chamber at the Space Research Institute of the Russian Academy of Sciences. This calibration enabled the determination of measurement accuracy for $CO_2$, $CH_4$, and $H_2O$ concentrations. Additionally, we conducted simultaneous measurements over 6 hours using LI-COR LI-7200 and LI-7700 analyzers at a grass field site in the Moscow region.

### 3.1. E-CAHORS Prototype Calibration in the Vacuum Chamber

While the TDLAS technique is assumed to be calibration-free, the sWMS technique requires calibration based on a set of known concentration points. Thus, to convert the E-CAHORS instrument measurement in sWMS regime into ppm units, this spectrometer was calibrated in the Space Research Institute RAS vacuum chamber. We mixed nitrogen, carbon dioxide, methane, and water vapour in different proportions to produce several concentrations of $CH_4$, $CO_2$, and $H_2O$ in a nitrogen atmosphere.

The spectra, as well as the calibration curves, are presented in Figure 5. Within the operational range of concentrations, the calibration curves exhibit a linear response to changes in gas concentration, with linear regression coefficients of determination ($R^2$) being close to 0.999 in all cases.

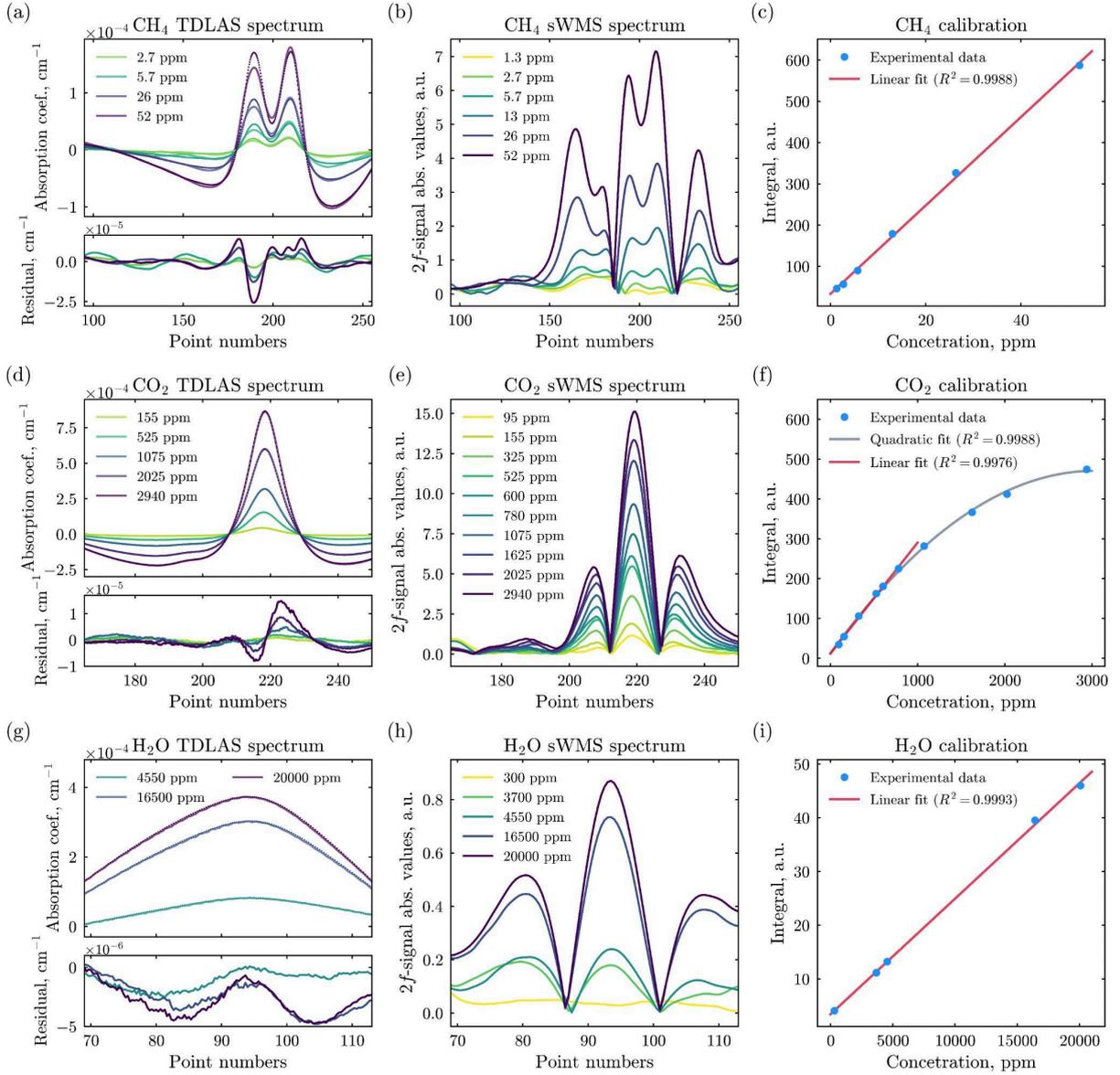

**Figure 5.** First column: baseline-corrected absorption coefficient spectra measured in the TDLAS regime and processed using the orthogonal polynomial method, along with the residuals between the measured and synthetic spectra at different concentrations for **(a)** methane, **(d)** carbon dioxide, and **(g)** water vapor. Second column: second harmonic signal $I_{d2}$ (2$f$-signal) module at different concentrations of **(b)** methane, **(e)** carbon dioxide, and **(h)** water vapor. Third column: calibration curves of the E-CAHORS instrument in sWMS regime for **(c)** methane, **(f)** carbon dioxide, and **(i)** water vapor, showing experimentally obtained values (blue dots), linear approximations (red lines), and quadratic approximation (grey line).

The E-CAHORS instrument installed inside the test vacuum chamber of the Space Research Institute RAS is shown in Figure 6(c).

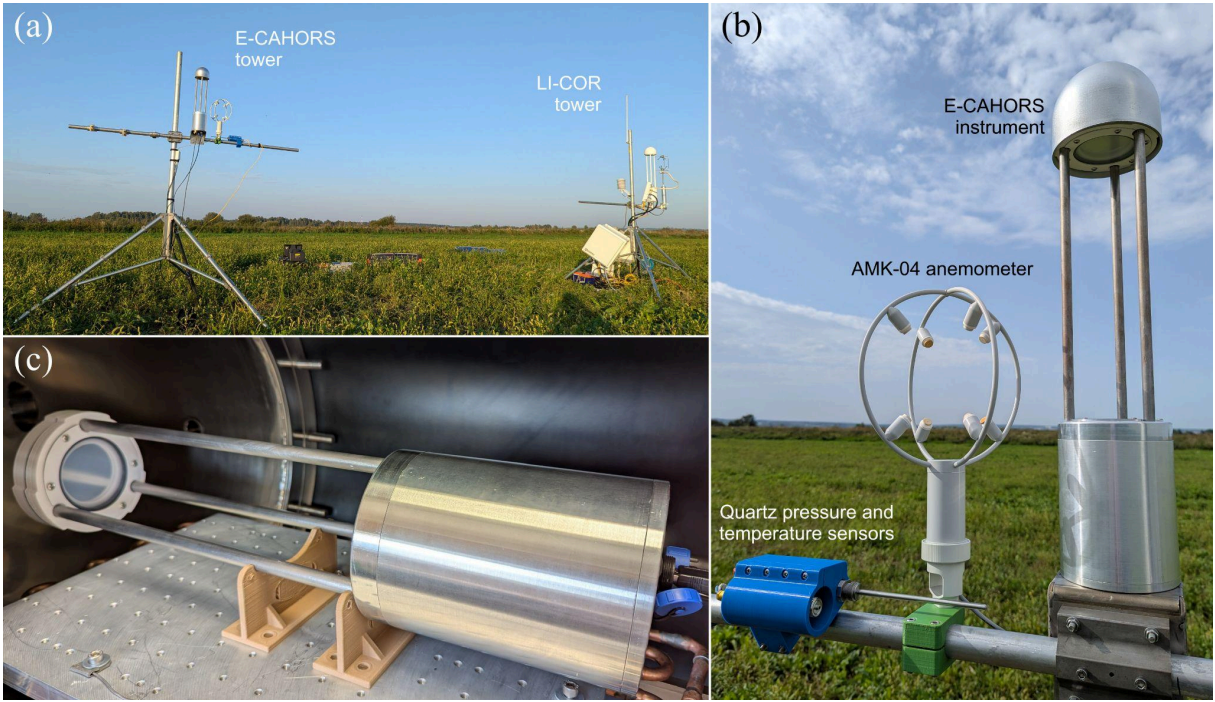

**Figure 6. (a)** Observation masts employed for parallel measurements of greenhouse gases in a sparsely urbanized area of the Moscow region: the mast equipped with the E-CAHORS instrument described in this study (left) and the mast fitted with two LI-COR gas analyzers (right). **(b)** A closer view of the E-CAHORS instrument, mounted on the mast together with the AMK-04 anemometer and quartz pressure and temperature sensors. **(c)** The E-CAHORS instrument in the vacuum chamber.

### 3.2. Instrument Precision Characterisation

The stability and precision of the instrument were characterized using Allan-Werle deviation analysis while the instrument was in the vacuum chamber filled with air at 1 atm and 298 K [45,46]. The study was performed for both TDLAS and sWMS regimes of operation, as demonstrated in Figure 7.

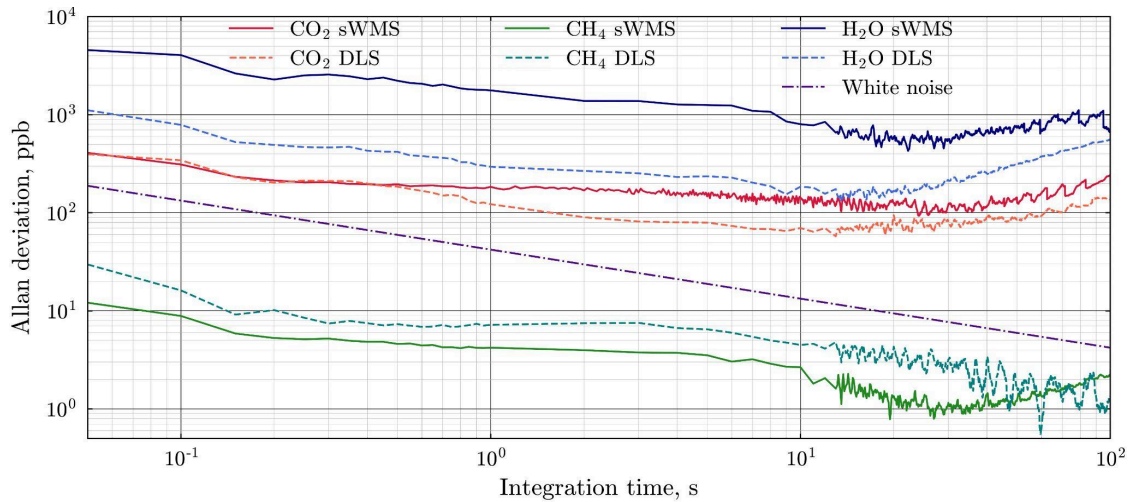

**Figure 7.** Allan-Werle deviation analysis in TDLAS and sWMS regimes of operation for continuous measurement of 470 ppm carbon dioxide, 2.2 ppm methane, and 3033 ppm water vapor. TDLAS results are shown with dashed curves, and sWMS results with solid curves. The color scheme: $CO_2$ in red tones, $CH_4$ in green, and $H_2O$ in blue.

We observe slightly different performance of sWMS compared to TDLAS at a sampling rate of 10 Hz: the lowest measurement deviations were 0.34 ppm in TDLAS and 0.31 ppm in sWMS for $CO_2$, 16.16 ppb in TDLAS and 8.87 ppb in sWMS for $CH_4$, and 0.79 ppm in TDLAS and 4.06 ppm in sWMS for $H_2O$. The best precision for TDLAS mode is presented in Table 1.

### 3.3. Field Measurements

The E-CAHORS instrument was set up on the mast in conjunction with an ultrasonic anemometer AMK-04 (Sibanalitpribor, Russia) with a sampling rate of 80 Hz [47,48] and quartz pressure and temperature sensors to test the performance of the developed gas analyzer in measuring $CO_2$, $CH_4$, and $H_2O$ mixing ratios. To validate the observed results, a LI-COR LI-7200 and LI-7700 were set up, along with a 3D-anemometer Gill R3-50 (Gill Instruments, UK), on the second mast. At the start of the campaign, the instruments were calibrated on a single measurement to ensure consistent mixing-ratio retrievals across LI-COR and E-CAHORS. The experimental site, located in the Moscow region near Dmitrov city, is shown in Figure 6(a). The measurements were conducted on a grassy field on a clear night from 22:30 to 0:00 (UTC+3) on August 29 and on a sunny day from 12:00 to 16:30 (UTC+3) on August 30, with a maximum wind speed of 12.8 m/s, a temperature range of 19.6-29.7 ℃, and an atmospheric pressure range of 995.46–997.22 mbar. The height of the E-CAHORS and LI-COR masts was 2 m, with a separation of 6.8 m between them. A closer view of the E-CAHORS instrument, mounted on the mast together with the anemometer and sensors used in this study, is presented in Figure 6(b).

As shown in Figure 8, both measurement methods produced similar trends in the concentrations of carbon dioxide, water vapor, and methane compared to the LI-COR instruments. However, the noise level in the TDLAS measurements was noticeably lower than in those obtained with the sWMS method.

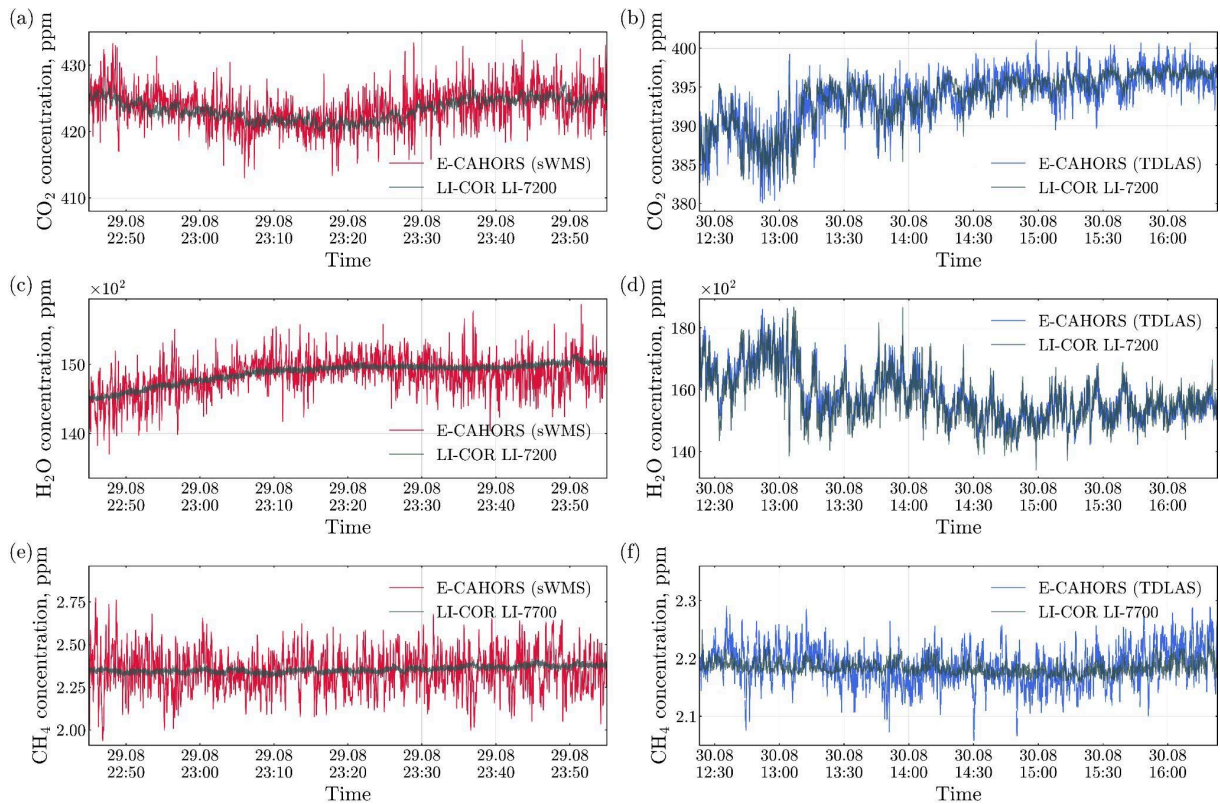

**Figure 8.** Comparison of atmospheric **(a,b)** $CO_2$ and **(c,d)** $H_2O$ concentration measurements obtained with the E-CAHORS instrument and the LI-COR LI-7200 analyzer, as well as **(e,f)** $CH_4$ concentration measurements from E-CAHORS and the LI-COR LI-7700 analyzer. In the left column, the E-CAHORS data recorded in the sWMS mode on the night of August 29 are shown in red, while in the right column, the E-CAHORS data recorded in the TDLAS mode during the day on August 30

are shown in blue. LI-COR data are shown in black. The distance between the instruments during field measurements was 6.8 meters.

Figure 9 presents the linear correlations of the measured concentrations of carbon dioxide, water vapor, and methane. The results show a strong agreement between E-CAHORS and LI-COR LI-7200 for $CO_2$ and $H_2O$, and a satisfactory correlation between E-CAHORS and LI-COR LI-7700 for $CH_4$. It should also be noted that, given the 6.8 m distance between the instruments, perfect agreement between the results should not be expected.

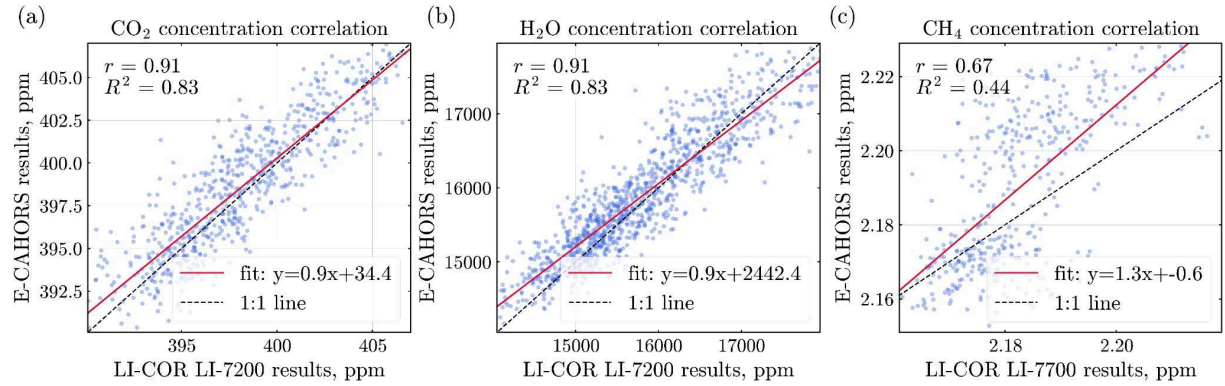

**Figure 9.** Linear correlation of **(a)** $CO_2$, **(b)** $H_2O$, and **(c)** $CH_4$ concentrations obtained with the E-CAHORS instrument (TDLAS method) and LI-COR devices. Data was averaged into intervals of 1 second to observe long trends.

*3.4. Discussion*

Based on the field evaluation of both measurement techniques implemented in the E-CAHORS instrument, it can be concluded that the TDLAS method demonstrated greater stability compared to sWMS and provided gas concentration measurements closely matching those from the LI-COR analyzers. The primary source of the observed difference between the two methods was the PTFE film used in the instrument's design to protect the spherical mirror. The film oscillated under strong wind conditions, introducing noise into the signal. This effect was less pronounced in the TDLAS mode because of signal normalization, but significantly affected the sWMS method, particularly for the relatively weak methane absorption lines, where it distorted the 2*f*-signal integral. In future, we plan to improve the sWMS data processing by implementing spectral line fitting using model-based data instead of relying solely on the 2*f*-integral, as well as by introducing normalization of the raw signal to compensate for laser power variation across the scanned spectral range.

The primary source of noise in the methane channel was fringes from the tightly built laser modules, which consist of an ICL with the collimating lens and three-axis alignment mechanism. We observed that the back-reflections from the lens would couple into the laser and produce strong fringes, which heavily reduced the accuracy of methane retrieval. Therefore, in subsequent iterations, we intend to use laser modules pre-aligned with collimating optics by the manufacturer. To further improve the signal-to-noise ratio for both techniques, we also plan to update the FPGA firmware to increase the acquisition rate.

We note that the current study was focused on validating the concentration measurements and identifying the design elements that must be addressed before flux measurements. Consequently, a long-term flux-measurement campaign is planned as the next step, as its success relies on the implementation of the improvements mentioned above.

Nevertheless, the described instrument has several distinguishing features compared to currently available analogs, as presented in the Table 2. Firstly, E-CAHORS can simultaneously measure the concentrations of three greenhouse gases – $CO_2$, $CH_4$, and $H_2O$ – while being lightweight and compact. Secondly, the open-path configuration allows for real-time observation of concentration fluctuations without the need to account for sample delay, which is typically introduced by closed-path systems that require active air pumping. Finally, the 4-pass optical system allows for longer unmaintained periods of operation, as there is a surplus of optical power to spare even with the dust sediment, and there are no moving components.

**Table 2.** Comparison of the developed instrument prototype, commercially available analogues, and those reported in the literature, suitable for the eddy covariance method of flux measurements.

| Instrument | Gas species | Resolution | Method | Sample path | Wave-length | Sampling rate | Mass | Power consum-ption |
|---|---|---|---|---|---|---|---|---|
| E-CAHORS (This work) | $CH_4$ | 8.87 ppb @10 Hz | sWMS / TDLAS | Open-path four-pass cell – 135 cm optical path length | 3240 nm | Up to 20 Hz | 4.2 kg | 10 W |
| | $CO_2$ | 0.31 ppm @10 Hz | | | 2783 nm | | | |
| | $H_2O$ | 0.79 ppm @10 Hz | | | 2783 nm, 3240 nm | | | |
| LI-7500DS (LI-COR Biosciences, USA) | $CO_2$ | 0.11 ppm @10 Hz | NDIR | Open-path single-pass cell | – | Up to 20 Hz | 1.6 kg | 4 W |
| | $H_2O$ | 4.7 ppm @10 Hz | | | | | | |
| LI-7200RS (LI-COR Biosciences, USA) | $CO_2$ | 0.11 ppm @10 Hz | NDIR | Closed-path single-pass cell | – | Up to 20 Hz | 1.8 kg | 12 W |
| | $H_2O$ | 4.7 ppm @10 Hz | | | | | | |
| IRGASON (Campbell Scientific, USA) | $CO_2$ | 0.15 ppm @20 Hz | NDIR | Open-path – 15.37 cm optical path length | – | Up to 60 Hz | 6 kg | 5 W |
| | $H_2O$ | 6 ppm @20 Hz | | | | | | |
| Li M. et al. (China) | $CO_2$ | 0.13 ppm @10 Hz | TDLAS | Open-path Herriott cell – 20 m optical path length | 2004 nm | Up to 100 Hz | – | – |
| | $H_2O$ | 3.25 ppm @10 Hz | | Open-path single-pass cell – 15 cm optical path length | 1392 nm | | | |
| G2311-f (Picarro Inc., USA) | $CH_4$ | 3 ppb @10 Hz | CRDS | Closed-path | 1603 nm | Up to 10 Hz | 37.7 kg | 360 W |
| | $CO_2$ | 0.2 ppm @10 Hz | | | 1651 nm | | | |
| | $H_2O$ | 6 ppm @10 Hz | | | 1603 nm | | | |
| Gu M. et al. (China) | $CO_2$ | 0.68 ppm @10 Hz | WMS | Open-path Herriot cell – 42.5 cm base length and 6.8 m optical path length | 2004 nm | Up to 10 Hz | 9.2 kg | 40 W |
| | $H_2O$ | 5.98 ppm @10 Hz | | Open-path two-pass cell – 42.5 cm base length and 85 cm optical path length | 1392 nm | | | |
| LI-7700 (LI-COR Biosciences, USA) | $CH_4$ | 5 ppb @10 Hz | WMS | Open-path Herriot cell – 47 cm base length and 30 m optical path length | 1.65 μm | Up to 20 Hz | 8.4 kg | 8 W |
| HT8600 (Healthy Photon Co., China) | $CH_4$ | 5 ppb @10 Hz | WMS | Open-path Herriot cell – 0.5 m base length and 46 m optical path length | 3221.1 nm | 10 Hz | 10 kg | 30 W |

## 4. Conclusion

In this paper, we presented the design of a laser-based open-path mid-IR spectrometer for greenhouse gas measurements. The instrument can simultaneously monitor the three most abundant greenhouse gases – $CO_2$, $H_2O$, and $CH_4$, suitable for flux retrieval via the eddy-covariance method. Its performance was evaluated through calibrations with known gas mixtures, which demonstrated linear responses of 0.998 for $CO_2$, and 0.999 for $CH_4$ and $H_2O$. The instrument's stability was assessed using Allan-Werle deviation analysis, resulting in the precisions of 311 ppb for $CO_2$, 8.87 ppb for $CH_4$, and 788 ppb for $H_2O$ at a 10 Hz measurement rate.

We also compared different spectroscopic techniques commonly used in similar instruments, focusing on Wavelength Modulation Spectroscopy (WMS) and Tunable Diode Laser Absorption Spectroscopy (TDLAS). In particular, we tested a simplified version of WMS (sWMS) with triangular modulation and evaluated its applicability for field measurements. The results showed that sWMS provided better precision for $CH_4$, similar precision for $CO_2$, and slightly worse precision for $H_2O$ compared to TDLAS at 10 Hz. While sWMS provides the baseline free spectra, we found that the technique is less reliable in outdoor conditions.

Finally, a field campaign was conducted over a grass field alongside LI-COR 7200 and LI-COR LI-7700 instruments. The TDLAS mode of our instrument showed good agreement with the LI-COR measurements, confirming its potential for outdoor flux studies. However, several aspects of the design require improvement before the planned long-term flux measurement campaign, including increasing the data acquisition rate, developing a more advanced mathematical model for sWMS, and reducing optical back reflections from the collimating lens.

With this work, we show that advances in mid-IR photonics make it possible to build reliable instruments for outdoor use. We developed a compact 4.2 kg instrument with a height of 670 mm that integrates the functionality of two conventional commercial spectrometers. Making such instruments more affordable could help create broader networks of measurement stations, improving our understanding of how human activity affects the environment on local scales.

# Declarations

**Funding** Partial financial support was received from the Russian Foundation for Basic Research under Grant Agreement № 18-29-24204.

**Conflict of interest** The authors declare no conflict of interest.

**Author contributions** Conceptualization: MS, SZ; Methodology: VM, IG, SZ; Formal analysis and data curation: BP, IG, VM, OK; Investigation: BP, VM, IG, VK, SG; Software: BP, VM, IG, VK; Hardware: IG, VK, SZ, SG; Validation: VM, BP, IG, VK, GS, YL, IV; Writing – original draft preparation: VM, IG; Visualisation: VM, IG; Writing – review and editing: MS, OK, BP; Funding acquisition: AR; Supervision: VM. All authors read and approved the final manuscript.

# References

1. Glacken, C.J. *Traces on the Rhodian Shore: Nature and Culture in Western Thought from Ancient Times to the End of the Eighteenth Century*; 6. [pr.].; Univ. of Calif. Pr: Berkeley, 1996; ISBN 978-0-520-02367-3.
2. Fourier, J. Mémoire Sur Les Températures Du Globe Terrestre et Des Espaces Planétaires. In *Mémoires de l'Académie Royale des Sciences de l'Institut de France*; 1827; Vol. 7, pp. 570–604.
3. Christianson, G.E. *Historical Perspectives on Climate Change.* By James Rodger Fleming. New York: Oxford University Press, 1998. Xi + 194 Pp. Illustrations, Notes, Bibliography, Index. $45.00. *Environ. Hist.* **2000**, *5*, 577–578, doi:10.2307/3985602.
4. Tyndall, J. On the Transmission of Heat of Different Qualities through Gases of Different Kinds. *Proc. R. Inst.* **1859**, *3*, 155–158.
5. Callendar, G.S. The Artificial Production of Carbon Dioxide and Its Influence on Temperature. *Q. J. R. Meteorol. Soc.* **1938**, *64*, 223–240, doi:10.1002/qj.49706427503.
6. Burba, G.G.; Anderson, D.J. *A Brief Practical Guide to Eddy Covariance Flux Measurements: Principles and Workflow Examples for Scientific and Industrial Applications*; LI-COR Biosciences, 2010;
7. Schmid, H.P. Source Areas for Scalars and Scalar Fluxes. *Bound.-Layer Meteorol.* **1994**, *67*, 293–318, doi:10.1007/BF00713146.
8. *Eddy Covariance Method: For Scientific, Regulatory, and Commercial Applications*; Updated and Expanded 2022 Edition.; LI-COR Biosciences: Lincoln, Nebraska, 2022; ISBN 978-0-578-97714-0.
9. Burba, G.; Schmidt, A.; Scott, R.L.; Nakai, T.; Kathilankal, J.; Fratini, G.; Hanson, C.; Law, B.; McDermitt, D.K.; Eckles, R.; et al. Calculating $CO_2$ and $H_2O$ Eddy Covariance Fluxes from an Enclosed Gas Analyzer Using an Instantaneous Mixing Ratio. *Glob. Change Biol.* **2012**, *18*, 385–399, doi:10.1111/j.1365-2486.2011.02536.x.
10. Polonik, P.; Chan, W.S.; Billesbach, D.P.; Burba, G.; Li, J.; Nottrott, A.; Bogoev, I.; Conrad, B.; Biraud, S.C. Comparison of Gas Analyzers for Eddy Covariance: Effects of Analyzer Type and Spectral Corrections on Fluxes. *Agric. For. Meteorol.* **2019**, *272–273*, 128–142, doi:10.1016/j.agrformet.2019.02.010.
11. Burba, G.G.; McDERMITT, D.K.; Grelle, A.; Anderson, D.J.; Xu, L. Addressing the Influence of Instrument Surface Heat Exchange on the Measurements of $CO_2$ Flux from Open‑path Gas Analyzers. *Glob. Change Biol.* **2008**, *14*, 1854–1876, doi:10.1111/j.1365-2486.2008.01606.x.
12. Haslwanter, A.; Hammerle, A.; Wohlfahrt, G. Open-Path vs. Closed-Path Eddy Covariance Measurements of the Net Ecosystem Carbon Dioxide and Water Vapour Exchange: A Long-Term Perspective. *Agric. For. Meteorol.* **2009**, *149*, 291–302, doi:10.1016/j.agrformet.2008.08.011.
13. Detto, M.; Verfaillie, J.; Anderson, F.; Xu, L.; Baldocchi, D. Comparing Laser-Based Open- and Closed-Path Gas Analyzers to Measure Methane Fluxes Using the Eddy Covariance Method. *Agric. For. Meteorol.* **2011**, *151*, 1312–1324, doi:10.1016/j.agrformet.2011.05.014.
14. McDermitt, D.K.; Welles J. M.; Eckles R. D. *Effects of Temperature, Pressure and Water Vapor on Gas Phase Infrared Absorption by CO2*; 1st ed.; LI-COR Technical Publication, 1993; Vol. 116;.
15. Burba, G.G.; McDermitt, D.K.; Anderson, D.J.; Furtaw, M.D.; Eckles, R.D. Novel Design of an Enclosed $CO_2$/$H_2O$ Gas Analyser for Eddy Covariance Flux Measurements. *Tellus B Chem. Phys. Meteorol.* **2010**, *62*, 743, doi:10.1111/j.1600-0889.2010.00468.x.
16. Herriott, D.; Kogelnik, H.; Kompfner, R. Off-Axis Paths in Spherical Mirror Interferometers. *Appl. Opt.* **1964**, *3*, 523, doi:10.1364/AO.3.000523.

17. McDermitt, D.; Burba, G.; Xu, L.; Anderson, T.; Komissarov, A.; Riensche, B.; Schedlbauer, J.; Starr, G.; Zona, D.; Oechel, W.; et al. A New Low-Power, Open-Path Instrument for Measuring Methane Flux by Eddy Covariance. *Appl. Phys. B* **2011**, *102*, 391–405, doi:10.1007/s00340-010-4307-0.
18. Olson, M.L.; Grieble, D.L.; Griffiths, P.R. Second Derivative Tunable Diode Laser Spectrometry for Line Profile Determination I. Theory. *Appl. Spectrosc.* **1980**, *34*, 50–56, doi:10.1366/0003702804731014.
19. Li, M.; Kan, R.; He, Y.; Liu, J.; Xu, Z.; Chen, B.; Yao, L.; Ruan, J.; Xia, H.; Deng, H.; et al. Development of a Laser Gas Analyzer for Fast CO2 and H2O Flux Measurements Utilizing Derivative Absorption Spectroscopy at a 100 Hz Data Rate. *Sensors* **2021**, *21*, 3392, doi:10.3390/s21103392.
20. Gu, M.; Chen, J.; Mei, J.; Tan, T.; Wang, G.; Liu, K.; Liu, G.; Gao, X. Open-Path Anti-Pollution Multi-Pass Cell-Based TDLAS Sensor for the Online Measurement of Atmospheric $H_2O$ and $CO_2$ Fluxes. *Opt. Express* **2022**, *30*, 43961, doi:10.1364/OE.474070.
21. Horst, T.W.; Vogt, R.; Oncley, S.P. Measurements of Flow Distortion within the IRGASON Integrated Sonic Anemometer and $CO_2/H_2O$ Gas Analyzer. *Bound.-Layer Meteorol.* **2016**, *160*, 1–15, doi:10.1007/s10546-015-0123-8.
22. Wang, K.; Kang, P.; Lu, Y.; Zheng, X.; Liu, M.; Lin, T.-J.; Butterbach-Bahl, K.; Wang, Y. An Open-Path Ammonia Analyzer for Eddy Covariance Flux Measurement. *Agric. For. Meteorol.* **2021**, *308–309*, 108570, doi:10.1016/j.agrformet.2021.108570.
23. Wang, K.; Wang, J.; Qu, Z.; Xu, W.; Wang, K.; Zhang, H.; Shen, J.; Kang, P.; Zhen, X.; Wang, Y.; et al. A Significant Diurnal Pattern of Ammonia Dry Deposition to a Cropland Is Detected by an Open-Path Quantum Cascade Laser-Based Eddy Covariance Instrument. *Atmos. Environ.* **2022**, *278*, 119070, doi:10.1016/j.atmosenv.2022.119070.
24. Baer, D.S.; Paul, J.B.; Gupta, M.; O'Keefe, A. Sensitive Absorption Measurements in the Near-Infrared Region Using off-Axis Integrated-Cavity-Output Spectroscopy. *Appl. Phys. B Lasers Opt.* **2002**, *75*, 261–265, doi:10.1007/s00340-002-0971-z.
25. Tuzson, B.; Hiller, R.V.; Zeyer, K.; Eugster, W.; Neftel, A.; Ammann, C.; Emmenegger, L. Field Intercomparison of Two Optical Analyzers for CH4 Eddy Covariance Flux Measurements. *Atmospheric Meas. Tech.* **2010**, *3*, 1519–1531, doi:10.5194/amt-3-1519-2010.
26. Crosson, E.R. A Cavity Ring-down Analyzer for Measuring Atmospheric Levels of Methane, Carbon Dioxide, and Water Vapor. *Appl. Phys. B* **2008**, *92*, 403–408, doi:10.1007/s00340-008-3135-y.
27. Chernin, S.M.; Barskaya, E.G. Optical Multipass Matrix Systems. *Appl. Opt.* **1991**, *30*, 51, doi:10.1364/AO.30.000051.
28. Chernin, S.M. Multipass Annular Mirror System for Spectroscopic Studies in Shock Tubes. *J. Mod. Opt.* **2004**, *51*, 223–231, doi:10.1080/09500340408235266.
29. Voigt, W. Über das Gesetz der Intensitätsverteilung innerhalb der Linien eines Gasspektrums.
30. Gordon, I.E.; Rothman, L.S.; Hargreaves, R.J.; Hashemi, R.; Karlovets, E.V.; Skinner, F.M.; Conway, E.K.; Hill, C.; Kochanov, R.V.; Tan, Y.; et al. The HITRAN2020 Molecular Spectroscopic Database. *J. Quant. Spectrosc. Radiat. Transf.* **2022**, *277*, 107949, doi:10.1016/j.jqsrt.2021.107949.
31. Bjorklund, G.C. Frequency-Modulation Spectroscopy: A New Method for Measuring Weak Absorptions and Dispersions. *Opt. Lett.* **1980**, *5*, 15, doi:10.1364/OL.5.000015.
32. Werle, P. A Review of Recent Advances in Semiconductor Laser Based Gas Monitors. *Spectrochim. Acta. A. Mol. Biomol. Spectrosc.* **1998**, *54*, 197–236, doi:10.1016/S1386-1425(97)00227-8.
33. Lins, B.; Zinn, P.; Engelbrecht, R.; Schmauss, B. Simulation-Based Comparison of Noise Effects in Wavelength Modulation Spectroscopy and Direct Absorption TDLAS. *Appl. Phys. B* **2010**, *100*, 367–376, doi:10.1007/s00340-009-3881-5.
34. Gazizov, I.; Pinto, D.; Moser, H.; Sam, S.; Martín-Mateos, P.; O'Faolain, L.; Lendl, B. Absorption and Dispersion: In Search of a Versatile Spectroscopic Technique. *Sens. Actuators B Chem.* **2025**, 137688, doi:10.1016/j.snb.2025.137688.
35. Meshcherinov, V.; Gazizov, I.; Kazakov, V.; Spiridonov, M.; Lebedev, Y.; Vinogradov, I.; Gerasimov, M. Spectrometer to Explore Isotopologues of Lunar Volatiles on Luna-27 Lander. *Planet. Space Sci.* **2024**, *248*, 105935, doi:10.1016/j.pss.2024.105935.
36. Rodin, A.; Vinogradov, I.; Zenevich, S.; Spiridonov, M.; Gazizov, I.; Kazakov, V.; Meshcherinov, V.; Golovin, I.; Kozlova, T.; Lebedev, Y.; et al. Martian Multichannel Diode Laser Spectrometer (M-DLS) for In-Situ Atmospheric Composition Measurements on Mars Onboard ExoMars-2022 Landing Platform. *Appl. Sci.* **2020**, *10*, 8805, doi:10.3390/app10248805.
37. Meshcherinov, V.V.; Spiridonov, M.V.; Kazakov, V.A.; Rodin, A.V. Lidar-Based Remote Infrared Gas Sensor for Monitoring Anthropogenic Pollution: A Proof of Concept. *Quantum Electron.* **2020**, *50*, 1055–1062, doi:10.1070/QEL17398.
38. Meshcherinov, V.; Kazakov, V.; Spiridonov, M.; Suvorov, G.; Rodin, A. Lidar-Based Gas Analyzer for Remote Sensing of Atmospheric Methane. *Sens. Actuators B Chem.* **2025**, *424*, 136899, doi:10.1016/j.snb.2024.136899.

39. Stewart, G.; Johnstone, W.; Bain, J.; Ruxton, K.; Duffin, K. Recovery of Absolute Gas Absorption Line Shapes Using Tuneable Diode Laser Spectroscopy with Wavelength Modulation ¿ Part I: Theoretical Analysis. *J. Light. Technol.* **2011**, 5696726, doi:10.1109/JLT.2011.2107312.
40. Bain, J.R.P.; Johnstone, W.; Ruxton, K.; Stewart, G.; Lengden, M.; Duffin, K. Recovery of Absolute Gas Absorption Line Shapes Using Tunable Diode Laser Spectroscopy With Wavelength Modulation—Part 2: Experimental Investigation. *J. Light. Technol.* **2011**, *29*, 987–996, doi:10.1109/JLT.2011.2107729.
41. Andreev, S.N.; Nikolaev, I.V.; Ochkin, V.N.; Savinov, S.Y.; Spiridonov, M.V.; Tskhai, S.N. Frequency Modulation upon Nonstationary Heating of the p − n Junction in High-Sensitive Diode Laser Spectroscopy. *Quantum Electron.* **2007**, *37*, 399–404, doi:10.1070/QE2007v037n04ABEH013311.
42. Mironenko, V.R.; Kuritsyn, Y.A.; Liger, V.V.; Bolshov, M.A. Data Processing Algorithm for Diagnostics of Combustion Using Diode Laser Absorption Spectrometry. *Appl. Spectrosc.* **2018**, *72*, 199–208, doi:10.1177/0003702817732252.
43. Wang, Z.; Fu, P.; Chao, X. Baseline Reduction Algorithm for Direct Absorption Spectroscopy with Interference Features. *Meas. Sci. Technol.* **2020**, *31*, 035202, doi:10.1088/1361-6501/ab54e4.
44. Kochanov, R.V.; Gordon, I.E.; Rothman, L.S.; Wcisło, P.; Hill, C.; Wilzewski, J.S. HITRAN Application Programming Interface (HAPI): A Comprehensive Approach to Working with Spectroscopic Data. *J. Quant. Spectrosc. Radiat. Transf.* **2016**, *177*, 15–30, doi:10.1016/j.jqsrt.2016.03.005.
45. Werle, P.; Mücke, R.; Slemr, F. The Limits of Signal Averaging in Atmospheric Trace-Gas Monitoring by Tunable Diode-Laser Absorption Spectroscopy (TDLAS). *Appl. Phys. B Photophysics Laser Chem.* **1993**, *57*, 131–139, doi:10.1007/BF00425997.
46. Werle, P. Accuracy and Precision of Laser Spectrometers for Trace Gas Sensing in the Presence of Optical Fringes and Atmospheric Turbulence. *Appl. Phys. B* **2011**, *102*, 313–329, doi:10.1007/s00340-010-4165-9.
47. Azbukin, A.A.; Bogushevich, A.Ya.; Korol'kov, V.A.; Tikhomirov, A.A.; Shelevoi, V.D. A Field Version of the AMK-03 Automated Ultrasonic Meteorological Complex. *Russ. Meteorol. Hydrol.* **2009**, *34*, 133–136, doi:10.3103/S1068373909020113.
48. Korolkov, V.A.; Telminov, A.E.; Tikhomirov, A.A. Metrological Support of Ultrasonic Thermoanemometers for Measurement of Pulsation Properties of Meteorological Parameters. *Atmospheric Ocean. Opt.* **2016**, *29*, 95–102, doi:10.1134/S1024856016010115.